\documentclass[pra,twocolumn,showpacs,preprintnumbers,amsmath,amssymb]{revtex4}
\usepackage{mathrsfs}
\usepackage{bbm}
\usepackage{amsfonts}
\usepackage{tipa}

\usepackage{epsfig,graphicx}
\usepackage{amstext}
\usepackage{amsmath}
\usepackage{graphicx}
\usepackage{times}
\usepackage{graphicx}

\begin{document}


\title{Thermal quantum discord in the Heisenberg chain with impurity}
\author{Jia-Min Gong}
\email{jmgong@yeah.net}
\author{Zhan-Qiang Hui}
\address{School of Electronic Engineering, Xi'an University of Posts and
         Telecommunications, Xi'an 710121, China}

\begin{abstract}
We study thermal quantum discord (TQD) in the Heisenberg chain with
spin site or magnetic impurity. The former one of which may induce
inhomogeneous exchange interactions between the neighboring spins,
while the latter one modelling a spin chain with nonuniform magnetic
field. In contrast with one's traditional understanding, we found
that the spin impurity can be used to enhance the TQD greatly for
all the bipartition schemes of the chain, while the magnetic
impurity located on one spin can make the TQD between the other two
spins approaching its maximum 1 for the antiferromagnetic chain.
\end{abstract}

\maketitle

\section{Introduction}\label{sec:1}
The distinctive features of quantum mechanics qualifies one to carry
out many information processing tasks which cannot be done in a
classical way \cite{Ekert,dense,tele1,tele2,tele3,qcomm,transfer}.
The existence of quantum correlations in a system was considered to
be responsible for the advantage of this way of information
processing \cite{Horodecki}, and this makes the quantification and
understanding of quantum correlations a vital problem needs to be
solved.

For a long time, the study of quantum correlations are focused on
entanglement \cite{Horodecki}, which has been shown to be a precious
resource in quantum information processing (QIP), and entanglement
exists only in non-separable states. But recent studies revealed
that quantumness other than entanglement can also exists in
separable states \cite{Modi}. Particularly, there are quantum
algorithms which outperform their classical counterparts while with
vanishing or negligible entanglement \cite{dqc1}. It is assumed that
quantum discord (QD) \cite{Ollivier}, a more fundamental measure of
quantum correlation than that of entanglement, provides speedup for
this task.

Due to the role it played in QIP \cite{Dakic,Gumile,Madhok,Winter},
and its fundamentals in quantum mechanics \cite{Werlang,Huml,Hu2},
QD has become one of people's research focuses in recent years
\cite{Streltsov,luo1,luo2,fan1,fan2}. Particularly, as a natural
candidate for implementing QIP tasks, the spin-chain systems have
attracted researcher's great interests, and the behaviors of QD in
various spin chains were analyzed \cite{sp1,sp2,sp3,sp4,sp5,sp6}.
More importantly, it has been found that the QD can serve as an
efficient quantity for detecting critical points of quantum phase
transitions even at finite temperature \cite{qpt1,qpt2,Werlang},
while the entanglement cannot achieve the same feat.

Different from previous studies which focused on QD behaviors in the
spin chain with only homogeneous exchange interactions
\cite{sp1,sp2,sp3,sp4,sp5,sp6}, here we go one step further and
consider QD behaviors in the Heisenberg model with inhomogeneous
interactions. This model can be viewed as a spin chain with spin
site imperfection or impurity \cite{simpurity,simpurity2}, and the
strength of interactions between the impurity spin and its
neighboring spins can be different from that between the normal
spins. Here, we will show that while being considered to be an
unwanted effects traditionally, the spin impurity can also serve as
an efficient way for controlling QD. Particularly, states with
considerable amount of QD exists in a wide regime of the
spin-impurity-induced inhomogeneous interaction, and this shows the
positive side of this unwanted effects.

Besides spin impurity, we will also consider the effect of a
nonuniform magnetic field on QD in the Heisenberg model. This model
describes the situation in which a magnetic impurity is located on
one site of the chain, and the entanglement properties in a similar
but different model (i.e., \emph{XX} chain) has already been
discussed \cite{mimpurity}. Here, we will further show that if the
magnetic impurity is located on the spin which is being traced out,
then the QD between the other two spins can be enhanced
asymptotically to its maximal value 1 for the antiferromagnetic
chain.

The structure of the following text is organized as follows. In
Section \ref{sec:2}, we give a brief review of QD and its
quantification based on the discrepancy between two expressions of
mutual information extended from classical to quantum system. Then
in Sections \ref{sec:3} and \ref{sec:4}, we introduce the spin model
we considered and discuss QD behaviors under different system
parameters. Finally, we conclude this work with a summary of the
main finding in Section \ref{sec:5}.

\section{Basic formalism of QD}\label{sec:2}
We recall in this section some basic formalism for QD. Up to now,
many measures of QD have been proposed, and they can be classified
into two categories in general \cite{Modi}. The first category are
those based on the entropic quantities
\cite{Ollivier,qce1,qce2,qce3}, while the second category are
defined via the geometric approach based on different distance
measures \cite{qcg1,qcg2,qcg3}. We adopt in this work the original
definition of QD proposed by Ollivier and Zurek \cite{Ollivier}. If
one denotes the quantum mutual information for a bipartite density
matrix $\rho_{AB}$ as $I(A:B)=S(\rho_A)+S(\rho_B)-S(\rho_{AB})$,
then the QD can be defined as
\begin{equation}\label{eq1}
 D(\rho_{AB})=I(A:B)-J(\rho_{AB}),
\end{equation}
where $J(\rho_{AB})$ is the so-called classical correlation of the
following form \cite{Henderson}
\begin{equation}\label{eq2}
 J(\rho_{AB})=S(\rho_B)-\inf_{\{E_k^A\}}S(B|\{E_k^A\}),
\end{equation}
where $S(B|\{E_k^A\})=\sum_{k}p_{k}S(\rho_{B|k})$ represents the
average conditional entropy of the postmeasurement state
$\rho_{B|k}={\rm Tr}_{A}(E_k^A\rho_{AB})/p_k$, with
$p_k=\text{Tr}(E_k^A\rho_{AB})$ being the probability for obtaining
the measurement outcome $k$, and the positive operator valued
measurements are performed on $A$.

The QD defined above was considered to be potential resource for
many QIP tasks \cite{dqc1,Gumile}, but its closed expression can
only be obtained for certain special states
\cite{analy1,analy2,analy3,analy4,Cenlx,Chitambar}, and it has been
shown to be an impossible-to-solve problem for general quantum
states \cite{Girolami}. Our discussion in this paper deals with only
qubit states, thus we can resort to numerical methods, and in view
of the generally negligible improvement by doing minimization over
full POVMs \cite{Hamieh,Galve}, we restrict ourselves to the
projective measurements by choosing the measurement operators as
$\Pi_1^A=|k_1^A\rangle\langle k_1^A|$ and $\Pi_2^A=I-\Pi_1^A$, with
$k_1^A=\cos(\theta/2)|0\rangle+e^{i\phi}\sin(\theta/2)|1\rangle$.
Moreover, for the two-qubit \emph{X} state with elements
$\rho_{AB}^{22}=\rho_{AB}^{33}$, the infimum of the conditional
entropy in Eq. \eqref{eq2} can be derived as \cite{Hu3}
\begin{equation}\label{eq3}
 \inf_{\{\theta,\phi\}}S(B|\{E_k^A\})=H(\tau),
\end{equation}
with the variable $\tau$ in the Shannon entropy function $H(\tau)$
being given by
\begin{equation}\label{eq4}
 \tau=\frac{1-\sqrt{[1-2(\rho_{AB}^{11}+\rho_{AB}^{33})]^2+
      4(|\rho_{AB}^{14}|+|\rho_{AB}^{23}|)^2}}{2}.
\end{equation}
Thus both $J(\rho_{AB})$ and $D(\rho_{AB})$ can be derived
analytically for this class of states.

\section{The model}\label{sec:3}
We consider in this paper the three-qubit Heisenberg chain in the
thermodynamic limit with the imposition of the periodic boundary
condition. The first case we will discuss is the chain for which the
neighboring spins coupled with inhomogeneous strengths, and the
Hamiltonian can be written as
\begin{equation}\label{eq5}
 \hat{H}=J_1 (\vec{\sigma}_1\cdot\vec{\sigma}_2+
         \vec{\sigma}_3\cdot\vec{\sigma}_1)+
         J\vec{\sigma}_2\cdot\vec{\sigma}_3,
\end{equation}
where $\vec{\sigma}_n=(\sigma_n^x,\sigma_n^y,\sigma_n^z)$ is the
vector of Pauli matrices, while $J_1$ and $J$ are the coupling
strengths, and $\hbar=1$ in Eq. \eqref{eq5} is assumed.

The above model can be viewed as a ring with an spin impurity at
site $1$. When it is thermalized with an external reservoir at
temperature $T$, the canonical ensemble can be evaluated by the
following density matrix,
\begin{equation}\label{eq6}
 \rho(T)=Z^{-1}\exp(-\hat{H}/k_B T),
\end{equation}
with $Z=\text{Tr}[\exp(-\hat{H}/k_B T)]$ being the partition
function and $k_B$ the Boltzman's constant, which will be set to
unity in the following text. As $\rho(T)$ represents a thermal
state, the QD in this state is called the thermal quantum discord
(TQD) \cite{Werlang}.

The eigenvalues as well as the eigenvectors of the Hamiltonian
$\hat{H}$ in Eq. \eqref{eq5} can be derived analytically, for which
we denote them as $\epsilon_i$ and $|\Psi_i\rangle$ ($i=1,2,\cdots,
8$), respectively. They are given by $\epsilon_{1,2}=J-4J_1$,
$\epsilon_{3,4}=-3J$, $\epsilon_{5,6,7,8}=J+2J_1$, and
\begin{eqnarray}\label{eq7}
 && |\Psi_1\rangle = (|101\rangle+|110\rangle-2|011\rangle)/\sqrt{6},\nonumber\\
 && |\Psi_2\rangle = (|001\rangle+|010\rangle-2|100\rangle)/\sqrt{6},\nonumber\\
 && |\Psi_3\rangle = (|001\rangle-|010\rangle)/\sqrt{2},\nonumber\\
 && |\Psi_4\rangle = (|101\rangle-|110\rangle/\sqrt{2},\nonumber\\
 && |\Psi_5\rangle = (|001\rangle+|010\rangle+|100\rangle)/\sqrt{3},\nonumber\\
 && |\Psi_6\rangle = (|011\rangle+|101\rangle+|110\rangle)/\sqrt{3},\nonumber\\
 && |\Psi_7\rangle = |000\rangle, |\Psi_8\rangle = |111\rangle.
\end{eqnarray}
Then $\rho(T)=\sum_i\exp(-\epsilon_i /
T)|\Psi_i\rangle\langle\Psi_i|/\sum_i\exp(-\epsilon_i /T)$. In the
next section, we will compute TQD for the density matrices
$\rho_{12}(T)=\text{Tr}_3 \rho(T)$, $\rho_{23}(T)=\text{Tr}_1
\rho(T)$, and $\rho(T)$ with the bipartition $\{1-23\}$, and discuss
their behaviors for $T\geqslant 0$ by changing the system
parameters.

Besides spin impurity which induces inhomogeneous interactions
between the neighboring spins, we consider also the Heisenberg model
with a magnetic impurity \cite{mimpurity}, which modelling a spin
chain with nonuniform magnetic field. The corresponding Hamiltonian
is given by
\begin{equation}\label{eq8}
 \hat{H}=J(\vec{\sigma}_1\cdot\vec{\sigma}_2+
         \vec{\sigma}_2\cdot\vec{\sigma}_3+
         \vec{\sigma}_3\cdot\vec{\sigma}_1)+
         B\sigma_1^z,
\end{equation}
where $B$ is the nonuniform magnetic field along $z$-direction of
the first spin.

For this Hamiltonian, its eigenvalues are $\epsilon_{1,2}=3J\pm B$,
$\epsilon_{3,4}=-3J\pm B$, $\epsilon_{5,6}=\pm\eta_+$, and
$\epsilon_{7,8}=\pm\eta_-$. Its eigenvectors can also be obtained
analytically, which are of the following form
\begin{eqnarray}\label{eq9}
 && |\Psi_1\rangle = |000\rangle, |\Psi_2\rangle = |111\rangle,\nonumber\\
 && |\Psi_3\rangle = (|001\rangle-|010\rangle)/\sqrt{2},\nonumber\\
 && |\Psi_4\rangle = (|101\rangle-|110\rangle)/\sqrt{2},\nonumber\\
 && |\Psi_{5,6}\rangle = |001\rangle+|010\rangle-\frac{B+J\mp\eta_+}{2J}|100\rangle,\nonumber\\
 && |\Psi_{7,8}\rangle = |101\rangle+|110\rangle+\frac{B-J\pm\eta_-}{2J}|011\rangle.
\end{eqnarray}
where $\eta_\pm=\sqrt{B^2+9J^2\pm 2JB}$, and
$|\Psi_{5,6,7,8}\rangle$ are unnormalized.

\section{TQD in the Heisenberg chain with impurity}\label{sec:4}
The impurity plays an important role in condensed matter physics. We
discuss here effects of the spin and magnetic impurities on TQD, and
show that while generally being considered to be the unwanted
effects, proper engineering of the impurity can also be used to
enhance the TQD greatly.

\subsection{Spin impurity}
We consider first the case of the spin impurity on TQD. For the
reduced density matrices $\rho_{12}(T)$ and $\rho_{23}(T)$, as they
are \emph{X} states and satisfy the conditions required by Eq.
\eqref{eq3}, closed expressions of TQD can be obtained (we do not
list them here for concise of the presentation). For $\rho(T)$ with
the bipartition $\{1-23\}$, we compute the TQD numerically.
\begin{figure}
\centering
\resizebox{0.4\textwidth}{!}{%
\includegraphics{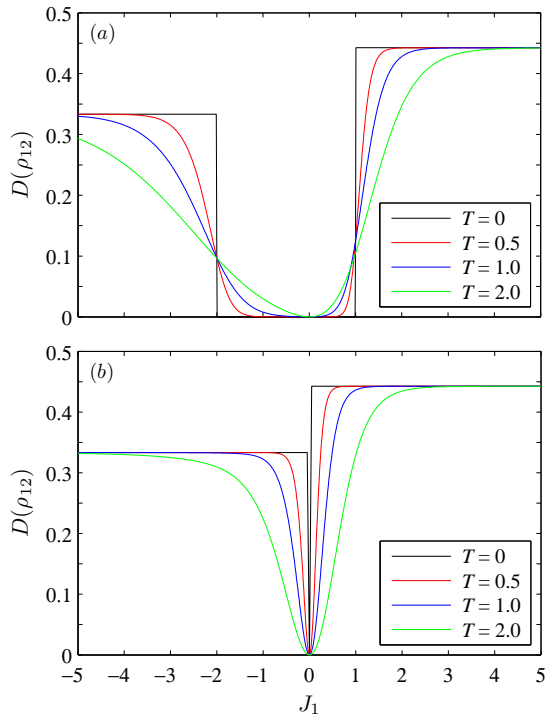}}
\caption{TQD $D(\rho_{12})$ between spins 1 and 2 versus $J_1$ with
         different scaled temperature $T$, and the parameter $J$ is chosen to
         be $J=1$ (a) and $J=-1$ (b), respectively.}
        \label{fig:1}
\end{figure}

When considering TQD between the impurity spin and the normal spin,
the $J_1$ dependence of $D(\rho_{12})$ with different $T$ are
displayed in Fig. \ref{fig:1}. First, at absolute zero temperature
with $J>0$, the ground states are the mixtures of
$|\Psi_{5,6,7,8}\rangle$ if $J_1<-2J$, $|\Psi_{3,4}\rangle$ if
$J_1\in(-2J,J)$, and $|\Psi_{1,2}\rangle$ if $J_1>J$, and the TQDs
are given by $1/3$, $0$, and $0.4425$, respectively. When $J<0$,
however, the ground states are the mixtures of
$|\Psi_{5,6,7,8}\rangle$ if $J_1<0$, which gives $D(\rho_{12})=1/3$;
and $|\Psi_{1,2}\rangle$ if $J_1>0$, which gives
$D(\rho_{12})=0.4425$.

For finite $T$, as can be seen from Fig. \ref{fig:1}, $D(\rho_{12})$
increases with increasing $|J_1|$, and when $J_1\rightarrow \infty$
and $-\infty$, we have $D(\rho_{12})=0.4425$ and $1/3$,
respectively. In order to obtain an intuition about the role of the
spin impurity played on enhancing TQD at finite $T$, we define the
critical $J_{1c}$ after which
$D(\rho_{12})|_{T=0}-D(\rho_{12})|_{T>0}<10^{-6}$. The numerical
fitting results performed in the region of $T\in(1,10)$ revealed
that if $J>0$, $J_{1c}$ satisfy the power law $J_{1c}\simeq 3.401
T+1.001$ when $J_1> J$, and $J_{1c}\simeq -7.450T-2.001$ when
$J_1<-2J$. Similarly, if $J<0$, we have $J_{1c}\simeq 3.401T-0.9846$
when $J_1>0$, and $J_{1c}\simeq -7.45T+1.999$ when $J_1<0$.

The above phenomena show that the TQD can be improved greatly
compared with that of the homogeneous Heisenberg chain (i.e.,
$J_1=J$), and the spin impurity can serve as an efficient way for
tuning TQD even at finite temperature. Moreover, it is worthwhile to
note that the TQD may be increased by increasing $T$ in the region
of $J_1\in(-2J,J)$ if $J>0$, and this peculiar phenomenon has also
been observed previously when studying entanglement
\cite{entangle1,entangle2}.
\begin{figure}
\centering
\resizebox{0.4\textwidth}{!}{%
\includegraphics{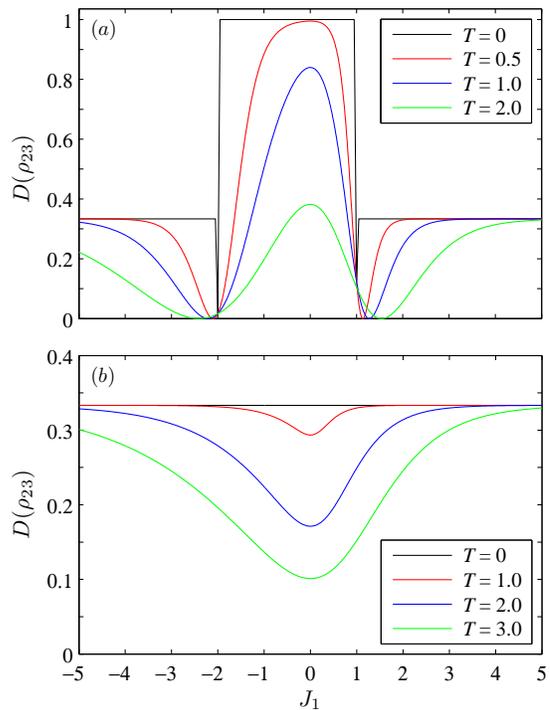}}
\caption{TQD $D(\rho_{23})$ between spins 2 and 3 versus $J_1$ with
         different scaled temperature $T$, and the parameter $J$ is chosen to
         be $J=1$ (a) and $J=-1$ (b), respectively.}
        \label{fig:2}
\end{figure}

When considering TQD between the normal spins 2 and 3 at absolute
zero temperature with $J>0$, we have $D(\rho_{23})=1/3$ in the
regions of $J_1<-2J$ and $J_1>J$, and $D(\rho_{23})=1$ in the region
$J_1\in(-2J,J)$. The latter case corresponds to mixtures of
$|\Psi_{3}\rangle$ and $|\Psi_{4}\rangle$, for which $\rho_{23}$
belongs to one of the Bell states and thus the result
$D(\rho_{23})=1$ is understandable \cite{Ollivier}. If $J<0$, the
ground state $\rho_{23}=(|01\rangle+|10\rangle)(\langle 01|+\langle
10|)/6+(|00\rangle\langle 00|+|11\rangle\langle 11|)/3$ in the full
region of $J_1$, and thus we always have $D(\rho_{23})=1/3$.

At finite temperature, the TQD $D(\rho_{23})$ is reduced with
increasing $T$ in a wide regime of $J_1$, and the exception appears
at the neighborhood of $J_1>J$ or $J_1<-2J$ for $J>0$ (see, Fig.
\ref{fig:2}). For any fixed $T$, $D(\rho_{23})$ can be enhanced
asymptotically to the steady-state value $1/3$ when
$|J_1|\rightarrow \infty$. In fact, for the same defined $J_{1c}$
above with $J>0$, we have $J_{1c}\simeq 4.245 T+1.001$ when $J_1>
J$, and $J_{1c}\simeq -8.144 T-2.001$ when $J_1<-2 J$. If $J<0$, we
have $J_{1c}\simeq 4.245 T-0.9992$ for $J_1> 0$, and $J_{1c}\simeq
-8.144 T+ 1.999$ for $J_1 < 0$. All these show again that
considerable enhancement of TQD can be achieved by tuning the
strength of the inhomogeneous exchange interaction $J_1$ moderately.
\begin{figure}
\centering
\resizebox{0.4\textwidth}{!}{%
\includegraphics{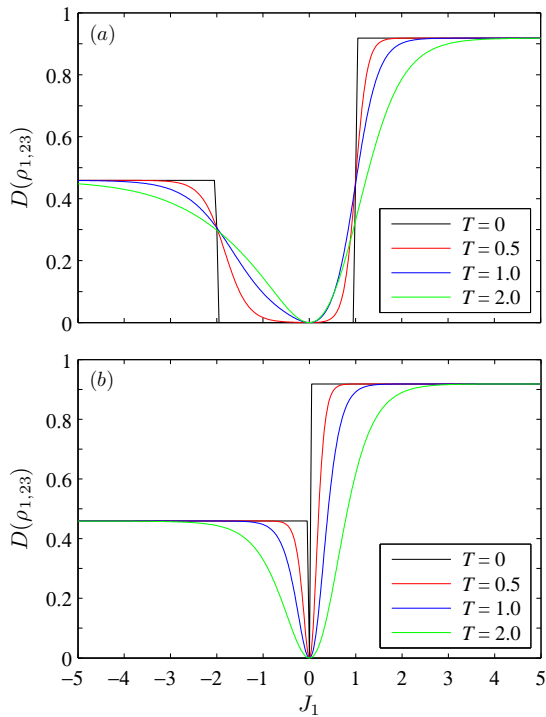}}
\caption{TQD $D(\rho_{1,23})$ between spin 1 and the spin pair
         $\{23\}$ versus $J_1$ with different scaled temperature $T$, and the
         parameter $J$ is chosen to be $J=1$ (a) and $J=-1$ (b),
         respectively.}
        \label{fig:3}
\end{figure}

We now discuss TQD for $\rho(T)$ with the bipartition $\{1-23\}$.
Here, we compute $D({\rho_{1,23}})$ numerically, and the
corresponding results are plotted in Fig. \ref{fig:3}. As the QD is
nonincreasing by tracing out one qubit \cite{Ollivier}, we have
$D({\rho_{1,23}})\geqslant D({\rho_{1,2}})$, which can be certified
by comparing Figs. \ref{fig:1} and \ref{fig:3}.

At absolute zero temperature with the addition of $J>0$, we have
$D({\rho_{1,23}})\simeq 0.9183$ when $J_1>J$,
$D({\rho_{1,23}})\simeq 0.4591$ when $J_1<-2J$, and
$D({\rho_{1,23}})=0$ when $J_1\in(-2J,J)$. If $J<0$, however, we
have $D({\rho_{1,23}})\simeq 0.9183$ $(0.4591)$ when $J_1>0$
($J_1<0$). At finite temperature $T$, the TQD $D({\rho_{1,23}})$
shows very similar behaviors with that of $D({\rho_{1,2}})$ (see,
Fig. \ref{fig:3}), i.e., it may be enhanced by increasing the
absolute value of $J_1$, and when $J_1\rightarrow \infty$
($-\infty$) it arrives at the asymptotic value $0.9183$ $(0.4591)$.
Moreover, the TQD is reduced by increasing temperature of the
reservoir except the special case of $J_1\in(-2J,J)$ and $J>0$.

\subsection{Magnetic impurity}
\begin{figure}
\centering
\resizebox{0.4\textwidth}{!}{%
\includegraphics{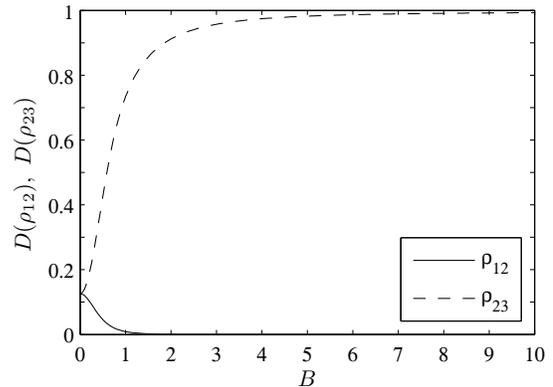}}
\caption{TQDs $D(\rho_{12})$ and $D(\rho_{23})$ versus the
         nonuniform magnetic field $B$ (along $z$-direction of spin 1) with
         $J=1$ and $T=0.25$.}
        \label{fig:4}
\end{figure}

For this situation, the interactions between the neighboring spins
are completely the same, and the magnetic impurity is assumed to be
along $z$-direction of the first spin.

We compute the TQD for different spin pairs numerically, and an
exemplified plot was displayed in Fig. \ref{fig:4} with $J=1$ and
$T=0.25$, from which one can see that with the increasing strength
of $B$, $D(\rho_{23})$ approaches to its maximum 1 asymptotically
[$D(\rho_{12})$ with $J>0$, or $D(\rho_{12})$ and $D(\rho_{23})$
with $J<0$ are always decreased with increasing $B$]. By tuning
strength of a nonuniform magnetic field located on one spin, one can
enhance the TQD between the other two spins. This phenomenon
reflects the remarkable nonlocal feature of
quantum mechanics.\\

\section{Summary}\label{sec:5}
To summarize, we have investigated properties of TQD in the
Heisenberg chain, which is assumed to be in thermal equilibrium with
a reservoir at temperature $T$. We considered the case that there
are spin site imperfection or magnetic impurity in the chain, and
discussed their influence on TQD between the chosen spin pairs. By
comparing its behaviors under different system parameters, we showed
that just as every coin has two sides, the unwanted effects of the
inhomogeneous exchange interaction induced by the spin impurity can
be used to improve the TQD greatly for all the bipartite states
considered. Moreover, we also showed that for the antiferromagnetic
Heisenberg chain with homogeneous exchange interactions, an magnetic
impurity along the $z$-direction of one spin can even be used to
make TQD between the other two spins approaching its maximal value
1, which is reminiscent of the nonlocal feature of quantum theory.

\newcommand{\PRL}{Phys. Rev. Lett. }
\newcommand{\RMP}{Rev. Mod. Phys. }
\newcommand{\PRA}{Phys. Rev. A }
\newcommand{\PRB}{Phys. Rev. B }
\newcommand{\PRE}{Phys. Rev. E }
\newcommand{\NJP}{New J. Phys. }
\newcommand{\JPA}{J. Phys. A }
\newcommand{\JPB}{J. Phys. B }
\newcommand{\OC}{Opt. Commun. }
\newcommand{\PLA}{Phys. Lett. A }
\newcommand{\EPJD}{Eur. Phys. J. D }
\newcommand{\NP}{Nat. Phys. }
\newcommand{\NC}{Nat. Commun. }
%

%

\end{document}